\documentclass[prb,aps,twocolumn,showpacs,floats,amssymb,amsfonts]{revtex4}

\usepackage{graphicx}
\usepackage{bm}
\newcommand{\sympt}[1]{\,\overline{{\mathrm{#1}}}\,}

\begin{document}

\title{
Heterovalent interlayers and interface states: an \textit{ab initio} study of
GaAs/Si/GaAs~(110) and (100) heterostructures
}

\author{M. Di Ventra}
\affiliation{
Department of Physics, University of California, San Diego, La Jolla, CA 92093-0319
}

\author{C. Berthod}
\affiliation{
DPMC, Universit\'e de Gen\`eve, 24 Quai Ernest-Ansermet, 1221 Gen\`eve 4,
Switzerland
}

\author{N. Binggeli}
\affiliation{
Abdus Salam International Center for Theoretical Physics and INFM DEMOCRITOS
National Simulation Center, Strada Costiera 11, 34014 Trieste, Italy
}

\date{\today}

\begin{abstract}

We have investigated {\it ab initio\/} the existence of localized states and
resonances in abrupt GaAs/Si/GaAs~(110)- and (100)-oriented heterostructures
incorporating 1 or 2 monolayers (MLs) of Si, as well as in the fully developed
Si/GaAs~(110) heterojunction. In (100)-oriented structures, we find both
valence- and conduction-band related near-band edge states localized at the
Si/GaAs interface. In the (110) systems, instead, interface states occur deeper
in the valence band; the highest valence-related resonances being about 1~eV
below the GaAs valence-band maximum. Using their characteristic bonding
properties and atomic character, we are able to follow the evolution of the
localized states and resonances from the fully developed Si/GaAs binary
junction to the ternary GaAs/Si/GaAs~(110) systems incorporating 2 or 1~ML of
Si. This approach also allows us to show the link between the interface states
of the (110) and (100) systems. Finally, the conditions for the existence of
localized states at the Si/GaAs~(110) interface are discussed based on a
Koster-Slater model developed for the interface-state problem.

\end{abstract}


\maketitle

\section{Introduction}

In recent years, a large amount of theoretical and experimental work has been
concentrated on heterostructures composed of ultrathin layers of a given
group-IV material deposited at III-V/III-V or II-VI/II-VI homo- or hetero-
junctions.\cite{alfonso} Interest in these systems has been stimulated by the
possibility of tuning the band discontinuity in heterojunctions, or inducing a
band discontinuity in homojunctions, thus allowing more flexibility in
designing novel devices.\cite{capasso} In particular, model
calculations,\cite{walter} supported by {\it ab initio\/}
computations,\cite{mariapert} have indicated that Si inserted as a bilayer at a
GaAs/AlAs~(100) heterojunction or at a GaAs/GaAs~(100) homojunction, can act as
a microscopic capacitor and induce substantial changes in the band
discontinuities. Although the interlayer morphology is still difficult to
control,\cite{franceschi,ploog} similar band-offset modifications have also
been reported experimentally.\cite{sorba,titti1,marsi,modesti} An issue,
however, which has remained largely unexplored concerning such engineered
interfaces is that of the electronic states induced by the interlayer. For any
practical application, it is indeed important to know whether the interlayer
will induce localized interface states within the bandgap that may degrade the
transport and optical properties of the junction.

Theoretical investigation of electronic states localized at interfaces between
heterovalent materials has a long history beginning with the pioneering work of
Baraff {\it et al.}\cite{baraf} on the Ge/GaAs~(100) interface. Since then,
several different heterojunctions have been investigated with special emphasis
on the interface-state
problem.\cite{baraf,pickett,pollman,saito,wang,laref,peressi} However, to our
knowledge, no theoretical study of the interface states has been performed so
far on the Si/GaAs heterojunction. Moreover, a complete understanding of the
mechanisms that control the formation of interface states and the conditions of
their existence is still lacking. As far as the problem of electronic states
induced by heterovalent interlayers is concerned, to our knowledge, studies in
this area have been limited to semiempirical tight-binding investigations on Ge
interlayers in GaAs, and of Si and GaAs interlayers in II-VI
semiconductors.\cite{saito,wang,laref} Tight-binding calculations, however,
neglect charge effects associated with the heterovalent nature of the
interlayer,\cite{saito,laref} and have yielded conflicting results (depending
on the parameters used) on the presence of localized states within the gap of
polar systems, such as the Ge/GaAs~(100) heterojunction.\cite{saito,pollman}

In the present work, we systematically investigate {\it ab initio\/} the
localized states and resonances induced by thin layers of Si in GaAs (110) and
(100) homojunctions, as well as in the fully developed Si/GaAs~(110)
heterojunction. Our results reveal interesting connections between the
formation of localized states and resonances in heterostructures possessing
different orientations and for interlayers of different thicknesses. A model is
also presented, which provides insight into the mechanisms of interface-state
formation; it explicitly links the existence of the heterojunction interface
states to some essential bulk band-structure features of the constituent
materials and some interface-specific bonding parameters.\cite{max} Since the
abrupt polar Si/GaAs~(100) interface is charged, and therefore gives rise to a
macroscopic electric field which complicates the study---and renders the fully
developed, abrupt (100) heterojunction thermodynamically
unstable\cite{walter}---we start by examining the interface-state spectrum of
the non-polar Si/GaAs~(110) heterojunction. We then study the electronic states
induced by 1 and 2~MLs of Si in the GaAs~(110) homojunction, and by 2~Si~MLs in
the GaAs~(100) homojunction. We discuss the relationship between these states
and their interface counterparts, and the conditions that govern the existence
of localized interface states.

\section{\textit{Ab initio} method}
\label{sect_method}

The interface states and interlayer-induced states are studied assuming
coherently strained Si layers on a GaAs substrate. The theoretical framework is
density functional theory (DFT) in the local density approximation (LDA), using
the Ceperley-Alder exchange-correlation functional.\cite{ceperley} We use
scalar relativistic Troullier-Martins pseudopotentials,\cite{troullier} in the
Kleinman-Bylander non-local form.\cite{kleinman} The valence electronic states
are expanded on a plane-wave basis set, using the supercell technique to model
the heterojunctions and heterstructures. We employ supercells including 12
monolayers (24 atoms) for the (110)-oriented structures and 16 monolayers (16
atoms) for the (100) structures. The supercell calculations are performed with
a kinetic-energy cutoff of 20 Ry and a $(6,6,2)$ Monkhorst-Pack $k$-point
mesh.\cite{monk} For the bulk calculations we use a 40-Ry cutoff and a
$(6,6,6)$ mesh. The calculated equilibrium lattice constants of bulk GaAs and
Si are 10.47 a.u. and 10.24 a.u., respectively. The resulting lattice mismatch
is $\sim2\%$, to be compared with the experimental value of
$\sim3.5\%$.\cite{lan} The theoretical values of the elastic constants of Si
are given in Table~\ref{tab:sigas}; they compare well with the experimental
values (also reported in the table), and with previous
calculations.\cite{peress94} Using the calculated values of the elastic
constants, and based on macroscopic elasticity theory (MET),\cite{van} the
theoretical lattice constant of Si on GaAs along the growth direction is 10.13
a.u. [10.08 a.u.] for the (110)-oriented [(100)-oriented] structures. In the
present work, for the starting (unrelaxed) Si/GaAs heterojunction and
GaAs/Si/GaAs heterostructures, we have used the MET result for the Si--Si
interlayer spacing, and, for the Si--GaAs interplanar distance at the
interface, we have used the average between the MET interlayer spacings in the
two bulks.

\begin{table}[tb]
\caption{\label{tab:sigas}
Theoretical equilibrium lattice parameter and elastic constants of Si used in
this work. Experimental values are reported in square brackets (from
Ref.~\onlinecite{lan}).
}
\begin{ruledtabular}
\begin{tabular}{lcccc}
   & $a_0$ (a.u.) & $c_{11}$ (Mbar) & $c_{12}$ (Mbar)& $c_{44}$ (Mbar) \\ \hline
Si & 10.24 [10.27]& 1.62 [1.67]     & 0.56 [0.65]    & 0.73 [0.80]
\end{tabular}
\end{ruledtabular}
\end{table}

The supercell electronic states have been classified according to their degree
of localization at the interface (or at the interlayer location) and their
atomic-like nature. We have considered for the interface-state spectrum only
those states whose probability density has a much higher value at the inversion
layer than in the bulk regions. Such states may be either truly localized
states, which decay exponentially in the bulk materials, or resonances.
Localized states are normally found in common gaps of the projected band
structures (PBS's) of the bulk materials,\cite{notesym} while resonances are
degenerate with Bloch states of one or both bulks. It is therefore convenient
to represent the interface states on a common energy scale with the PBS's of
the bulk materials. The alignment of the bulk band structures was performed
using the calculated valence-band offset (VBO) obtained with the technique
outlined in Ref.~\onlinecite{baldo}. The supercell interface-state spectrum has
then been aligned with respect to the top valence-band edge of one of the bulk
materials using the calculated shift of the
macroscopic-average\cite{baldo,maria} electrostatic potential in the
corresponding bulk region of the supercell relative to the mean value of the
electrostatic potential in the supercell.

In the case of the (110)-oriented structures, atomic relaxation was found to
have a relatively minor influence on the interface-state spectra. The choice of
the average bulk Si and GaAs MET spacings for the Si--GaAs interlayer distance
is a good starting point for these structures: the error with respect to the
relaxed interlayer distance is less than 1\%. Atomic relaxation has normally
two effects. It changes: ({\it i\/}) the band offset, i.e., the relative
position of the bulk PBS's and ({\it ii\/}) the relative position of the
interface states with respect to the bulk PBS. The first effect vanishes (by
symmetry) for the GaAs/Si/GaAs~(110) heterostructures, and is negligible for
the Si/GaAs~(110) heterojunction ($\Delta V \approx 0.02$~eV). The deviations
({\it ii\/}), instead, can be as large as 0.1~eV for the Si/GaAs~(110)
heterojunction. However, this is of the same order of magnitude as the
many-body and spin-orbit effects,\cite{maria} not included in our calculations.
Therefore, for the (110) structures we will consider in the following the ideal
unrelaxed geometries only.

Larger relaxation effects were found, instead, in the (100)-oriented
GaAs/2\,Si\,MLs/GaAs heterostructure. The As--Si (Ga--Si) interplanar distance
increases (decreases) by 7.5\% (0.5\%) and the Si--Si interlayer spacing
decreases by 2.4\%, with respect to the unrelaxed MET geometry. This results in
changes of several tenths of an eV in the energy position of the localized
states and resonances of the GaAs/2\,Si\,MLs/GaAs~(100) heterojunction. In the
following, all results presented for the GaAs/2\,Si\,MLs/GaAs~(100) system will
refer to the fully relaxed geometry.

We have also checked the effect of the supercell size using a 32-atom cell for
the Si/GaAs~(110) heterojunction. We found variations in the energy position of
the electronic states of at most 0.05~eV. We can therefore conclude that the
energy position of the electronic states in our (100) and (110)
heterostructures is given with an overall uncertainty of the order of 0.1~eV.

\section{Si/GaAs~(110) interface states}

\begin{figure}[tb]
\includegraphics[width=7cm]{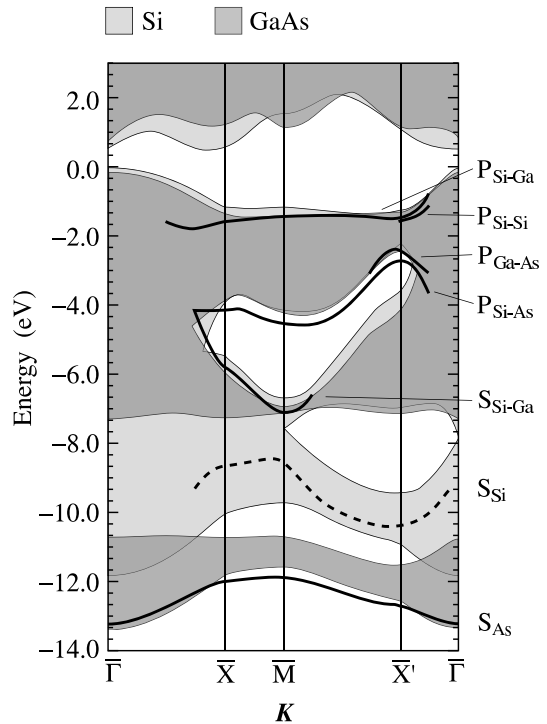}
\caption{\label{stateint}
Dispersion of the interface states (thick lines) of the Si/GaAs~(110)
heterojunction relative to the PBS of the Si and GaAs bulk materials (hatched).
The dashed line is a resonant state localized on the first Si bilayer from the
interface (see text).
}
\end{figure}

Following the procedure outlined in the previous section, we have calculated
the interface states of the Si/GaAs~(110) heterojunction. The result is
displayed in Fig.~\ref{stateint}, where we have represented the dispersion of
the states along the principal lines of the irreducible wedge of the
two-dimensional Brillouin zone (2DBZ), together with the two bulk PBS's. The
calculated VBO of the Si/GaAs~(110) heterojunction (without spin-orbit and
self-energy corrections) used for the alignment of the bulk PBS's is 0.14~eV
(Si higher). The high-symmetry points of the 2DBZ are $\sympt{\Gamma} = (0,0)$,
$\sympt{X} = (\sqrt{2},0)$, $\sympt{M} = (\sqrt{2},1)$, and $\sympt{X}' =
(0,1)$, in units of $\pi/a$ ($a$ is the GaAs lattice constant).

The isolated (110) interface has a $\sigma_{yz}$ reflection
symmetry\cite{koster} ($z$ is the growth direction and $y$ is the direction
perpendicular to the Si--Si or Ga--As nearest-neighbor zigzag chains in the
(110) atomic planes). However, since we have two interfaces per supercell an
additional (fictitious) reflection symmetry induces non-degenerate pairs of
symmetric and antisymmetric (by reflection) states in the quasi 2DBZ of the
supercell (the symmetry group of the supercell is $C_{2v}$). With the supercell
employed, though, the energy difference between the above pairs is about
$50$~meV throughout the BZ, thus indicating a small interaction between the two
interfaces.

\begin{figure}[tb]
\includegraphics[width=8.7cm]{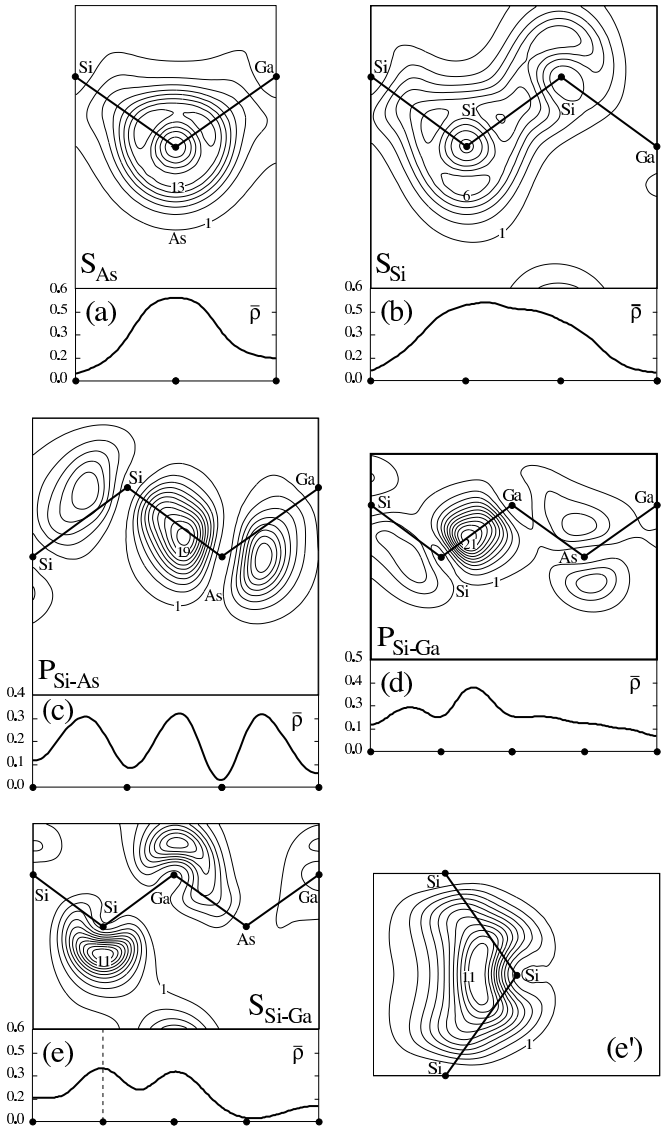}
\caption{\label{stateint1}
Contour plot of the probability density $\rho$ (in $e$/supercell) in the
$(\bar{1}10)$ plane and planar average of the probability density
$(\bar{\rho})$ along the [110] direction of selected interface states of the
Si/GaAs~(110) heterojunction. (a) As-derived $s$ state at $\sympt{X}'$. (b)
Si-derived $s$ state at $\sympt{X}'$. (c) Si and As derived $p$ state at
$\sympt{X}'$. (d) Si and Ga derived $p$ state at $\sympt{X}'$. (e) Si and Ga
derived $s$-$p$ states at $\sympt{M}$. (e') The contour plot of the (e) state
in the Si~(110) plane adjacent to the interface (dashed line in the (e) plot).
The probability density of the above states in atomic planes other than the
ones considered is a small fraction of the one shown.
}
\end{figure}

Beginning from the lowest energies, we find a complete band of localized states
below (or at the edge of) the bulk PBS of the $s$ valence band of GaAs; the
probability density of the state at the $\sympt{X}'$ point of the 2DBZ is shown
in Fig.~\ref{stateint1}(a). These states (S$_{\text{As}}$ in
Fig.~\ref{stateint}) derive from the $s$ states of the As atoms adjacent to the
interface. The localization of these states results from the more attractive
ionic potential of Si relative to Ga, which lowers the energy of the As-$s$
states at the interface with respect to those in bulk GaAs. This is consistent
with the relative strengths of the atomic potentials, which obey the inequality
Ga $<$ Si $<$ As, and leads to the following expected attractiveness of the
potential in the interfacial bonding regions: Si--Ga $<$ Si--Si $<$ Ga--As $<$
Si--As. We note that similar localized S$_{\text{As}}$ states were reported by
Pickett {\it et al.}\cite{pickett} for the Ge/GaAs~(110) interface.

Moving upwards in energy we find a second band of states which have a maximum
probability density on the first Si bilayer from the interface (S$_{\text{Si}}$
states, dashed line in Fig.~\ref{stateint}). These states are in the energy
region of the Si PBS only, and show a small coupling with the Si continuum.
Inspection of the probability density of these states, shown in
Fig.~\ref{stateint1}(b), indicates that they derive predominantly from the
atomic $s$-states of Si atoms in the second and first layer from the interface.
The authors of Ref.~\onlinecite{pickett} have found analogous resonant states
in the case of the Ge/GaAs~(110) interface.

The band in the $-7$ to $-4$~eV energy range along the $\sympt{X}$--$\sympt{M}$
line corresponds to states related to Si and Ga atoms at the interface (we
called them S$_{\text{Si-Ga}}$ states). The probability density of one of these
states at $\sympt{M}$ is displayed in two different planes, perpendicular and
parallel to the interface, in Figs.~\ref{stateint1}(e) and (e'), respectively.
Such states contain a large $s$ component, and also a non-negligible $p$
contribution, of the Si and Ga atoms at the interface, and exhibit a strong
antibonding character. They are pushed away from the Si--Ga bonding region into
the more attractive Si--Si and Ga--As antibonding regions. We note that at the
Ge/GaAs~(110) interface, the corresponding states were found to have a dominant
contribution from the Ga atom,\cite{pickett} which is not the case here. We
attribute this mainly to strain. Strain pushes the Si $s$ states upwards in
energy, giving rise to a stronger interaction with the Ga states, and hence to
a more significant contribution of the group-IV atom to the antibonding-like
interface state (this will become more clear from the study of the ultrathin
Si~(110) layers in the following sections).

The states of the next two bands that extend in the
$\sympt{X}$--$\sympt{M}$--$\sympt{X}'$ region (P$_{\text{Si-As}}$ and
P$_{\text{Si-Ga}}$) originate from $p$ states of the Si atoms at the interface
and from $p$ states of the interface As and Ga atoms, respectively (see
Fig.~\ref{stateint1}(c) and (d)), which form bonding orbitals across the
interface. For the localization of the P$_{\text{Si-As}}$ states, the same
considerations as for the S$_{\text{As}}$ states are valid. As far as the
P$_{\text{Si-Ga}}$ states are concerned, we note that due to the repulsive
ionic potential in the Si--Ga bonding region (with respect to the potential in
the Ga--As or Si--Si regions), these states are pushed at the edge of (or at a
higher energy than) the GaAs PBS.

\begin{figure}[tb]
\includegraphics[width=8.7cm]{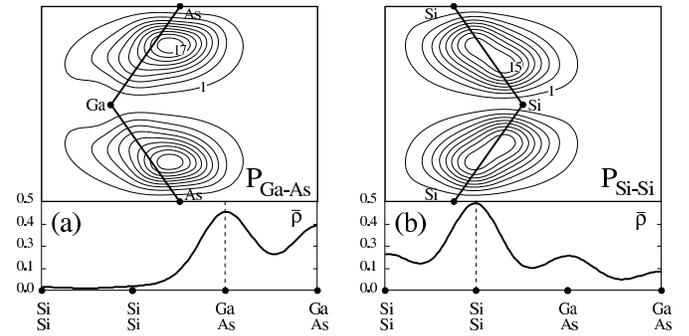}
\caption{\label{stateint2}
Contour plot of the probability density of two Si/GaAs~(110) interface states
in two different (110) planes containing Ga and As atoms (a) and Si atoms (b)
at the interface, and corresponding planar average $(\bar{\rho})$ along the
[110] direction. Both states are at the $\sympt{X}'$ point of the zone. The
dashed lines indicate the position of the planes where the contour plots were
drawn.
}
\end{figure}

The remaining two types of state (P$_{\text{Ga-As}}$ and P$_{\text{Si-Si}}$ in
Fig.~\ref{stateint}) are found only in the neighborhood of the $\sympt{X}'$
point, and have their maximum probability density either on the GaAs~(110)
plane (P$_{\text{Ga-As}}$) or on the Si~(110) plane (P$_{\text{Si-Si}}$)
adjacent to the interface (see Fig.~\ref{stateint2}). These states correspond
essentially to As--Ga (Si--Si) $p_x$ bonding orbitals within the
nearest-neighbor zigzag chains of the (110) plane. We note that the
$\sympt{X}'$ point corresponds to a zone-edge $\bm{K}$ vector perpendicular to
the chains. In a tight-binding description,\cite{harris} this gives rise to a
phase in the Bloch functions that leaves the sign of the $p_x$ orbitals of
second nearest neighbor atoms in the chain unchanged, but reverses the sign of
the atomic-like orbitals in adjacent chains within the (110) plane. This leads
to an antibonding configuration between the chains, but allows the formation of
$p_x$ nearest-neighbor bonding orbitals within the chain, as seen in
Fig.~\ref{stateint2}. Finally, the fact that the P$_{\text{Si-Si}}$
(P$_{\text{Ga-As}}$) states are lower (higher) in energy than the
P$_{\text{Si-Ga}}$ (P$_{\text{Si-As}}$) states can be explained by the more
attractive ionic potential of Si as compared to Ga.

\section{States induced by one Si~(110) monolayer in GaAs}

Let us return to the S$_{\text{Si}}$ states and also to the S$_{\text{Si-Ga}}$
states of the Si/GaAs~(110) heterojunction. Their formation and atomic
character can be better understood if only few layers of Si are introduced in
bulk GaAs. Indeed, the bulk PBS of Si is no longer present, which enhances the
localization of some of the Si induced states. We note that some of the states
of the Si/GaAs interface are bound to be present also in the ultra-thin layer
limit. In particular, all As and Ga-derived states cannot disappear since they
essentially originate from GaAs band-edge features. We will see here and in the
next section that actually most of the interface states of the fully developed
Si/GaAs~(110) heterojunction can be traced down to the ML limit.

\begin{figure}[tb]
\includegraphics[width=7cm]{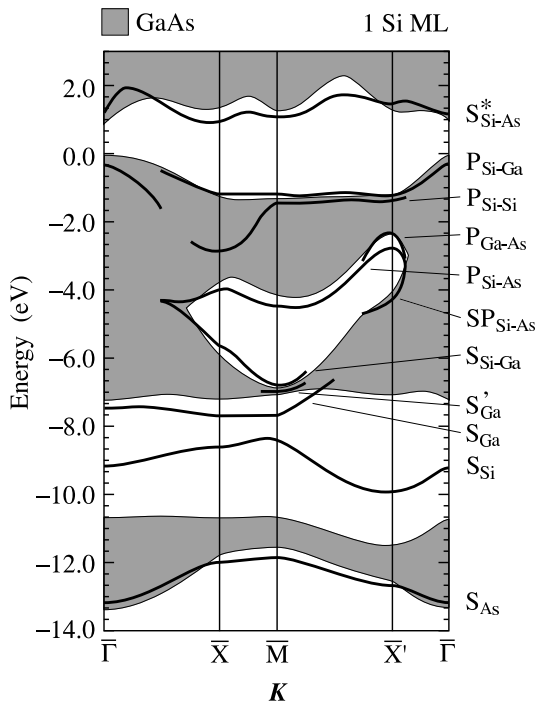}
\caption{\label{state1ML}
Dispersion of the localized states and resonances (thick lines) induced by 1~ML
of Si~(110) in GaAs relative to the GaAs bulk PBS.
}
\end{figure}

In Fig.~\ref{state1ML} we have plotted the states induced by 1~ML of Si in the
GaAs~(110) homojunction, together with the GaAs PBS. The layer-induced states
have been calculated at $k_z=0$. However, the miniband width along $k_z$ is
negligible given the thickness of the GaAs slab employed in the present study.
We have aligned the supercell states with the bulk GaAs PBS using the
calculated shift of the macroscopic average of the electrostatic potential in
GaAs relative to its mean value in the supercell. The group symmetry is
$C_{2v}$, and the space group is symmorphic. Thus, all points and lines in the
2DBZ belong to one-dimensional irreducible representations, and no additional
degeneracy is introduced by time reversal symmetry.

\begin{figure}[tb]
\includegraphics[width=8.7cm]{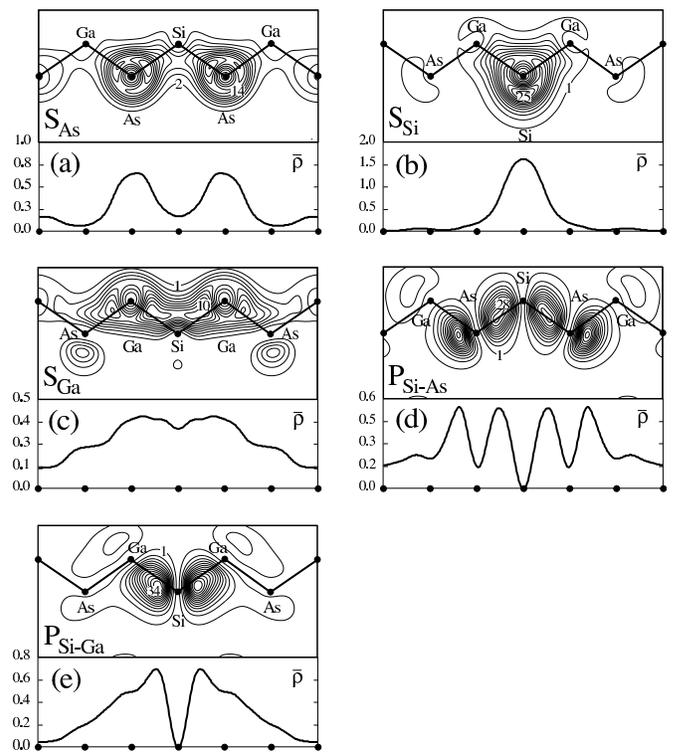}
\caption{\label{state1ML1}
Contour plot of the probability density $\rho$ in the $(\bar{1}10)$ plane and
planar average of the probability density $(\bar{\rho})$ along the [110]
direction of selected states of the GaAs/1\,Si\,ML/GaAs~(110) heterostructure.
(a) As-derived $s$ state at $\sympt{X}'$. (b) Si-derived $s$ state at
$\sympt{X}'$. (c) Ga-derived $s$ state at $\sympt{\Gamma}$. (d) Si and As
derived $p$-state at $\sympt{X}'$. (e) Si and Ga derived $p$-state at
$\sympt{X}'$. The probability density of the above states in atomic planes
other than the ones considered is a small fraction of the one shown.
}
\end{figure}

An S$_{\text{As}}$ band is still present in Fig.~\ref{state1ML} at
approximately the same energy as in the interface case. This band derives from
the $s$ states of the As atoms adjacent to the Si layer that form symmetric
states under the $\sigma_{xy}$ reflection (see Fig.~\ref{state1ML1}(a)). The
antisymmetric counterpart is found as a broad resonance (not shown in
Fig.~\ref{state1ML}) in the PBS of the bulk $s$ valence band of GaAs.
Conversely, a Si-related $s$ band appears at about $-9$~eV at $\sympt{\Gamma}$,
whose dispersion from $\sympt{\Gamma}$ to $\sympt{X}'$, to $\sympt{M}$, and to
$\sympt{X}$ is similar to that of the S$_{\text{Si}}$ interface state. Again,
this band corresponds to symmetric states with respect to $\sigma_{xy}$. The
probability density of one of these states (S$_{\text{Si}}$ at $\sympt{X}'$) is
displayed in Fig.~\ref{state1ML1}(b). The probability density is predominantly
on the Si atoms sitting on the anion sites in the continuation of the
zincblende lattice, but also reveals some bonding character with the
nearest-neighbor Ga and Si atoms. Considering the Si layer as a perturbation to
bulk GaAs,\cite{KoSl54,withkurt} these states separate from the GaAs $s$ band
and move up in energy, because of the repulsive on-site potential associated
with the Si $\rightarrow$ As substitutions. We note that because these
localized states have their probability density mostly on the anion sites,
within the Si~(110) layer at the interface, they are qualitatively different
from the Si bulk states present at similar energies in Fig.~\ref{stateint}; the
latter states derive from Si atomic orbitals with identical weight on the two
Si sublattice sites. The former states can thus give rise to Fano resonances
when they interact with the continuum of bulk Si states, in the fully developed
Si/GaAs~(110) heterojunction. The latter interaction generates bonding features
between the Si atoms sitting on anion sites, in the layer closest to GaAs, and
the Si atoms sitting on cation sites, in the neighboring layer on the Si side.
This, in turn, pushes the probability density of the S$_{\text{Si}}$ state
towards the second Si layer, when the heterojunction is formed.

\begin{figure}[tb]
\includegraphics[width=8.7cm]{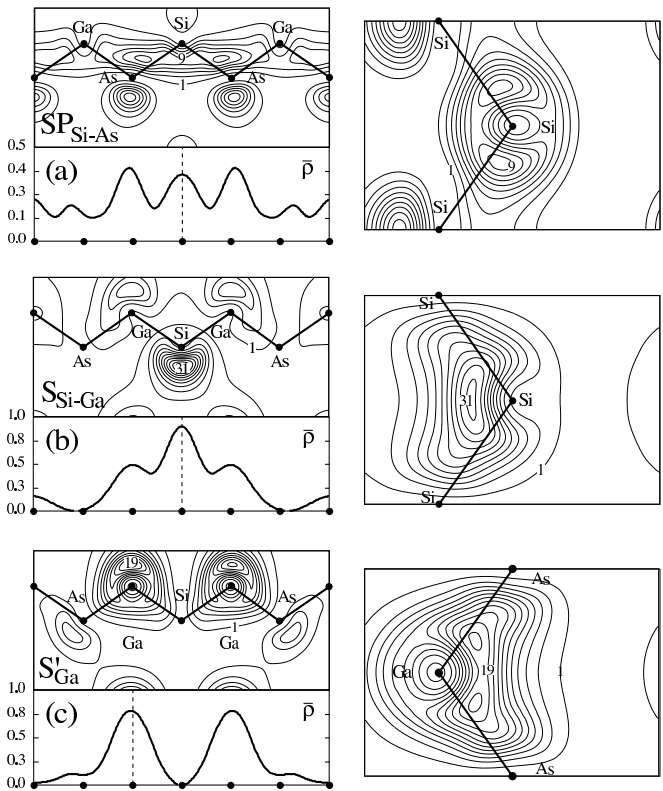}
\caption{\label{state1ML2}
Left: contour plot of the probability density $\rho$ in the $(\bar{1}10)$ plane
and planar average of the probability density $(\bar{\rho})$ along the [110]
direction of three states of the GaAs/1\,Si\,ML/GaAs~(110) heterostructure.
Right: contour plot of the same states in the (110) plane indicated by the
dashed line. (a) $sp$-Si and $p$-As derived state at $\sympt{X}'$. (b) Si and
Ga-derived $s$ state at $\sympt{M}$. (c) Ga-derived $s$ state at $\sympt{M}$.
The probability density of the above states in atomic planes other than the
ones considered is a small fraction of the one shown.
}
\end{figure}

In the neighborhood of the $\sympt{M}$ point, along the
$\sympt{\Gamma}$--$\sympt{X}$--$\sympt{M}$--$\sympt{X}'$ path in
Fig.~\ref{state1ML}, the S$_{\text{Si}}$ state couples with the symmetric state
we labeled S$_{\text{Ga}}$. The latter state derive mainly from the $s$ states
of the Ga atoms adjacent to the Si layer (its probability density at
$\sympt{\Gamma}$ is shown in Fig.~\ref{state1ML1}(c)). Consistent with the
ordering expected from the corresponding $s$ levels in the isolated atoms, the
energy positions of the localized $s$ states of As, Si, and Ga are found in
ascending order. The state labeled S$_{\text{Ga}}'$, observed at $\sympt{M}$,
is the antisymmetric counterpart of the S$_{\text{Ga}}$ state (see
Fig.~\ref{state1ML2}(c)), while the symmetric S$_{\text{Si-Ga}}$ state, found
in the lower part of the GaAs stomach gap, has mixed Si and Ga $s$ (with some
$p$) atomic character (see Fig.~\ref{state1ML2}(b)).

\begin{figure*}[tb]
\includegraphics[width=15cm]{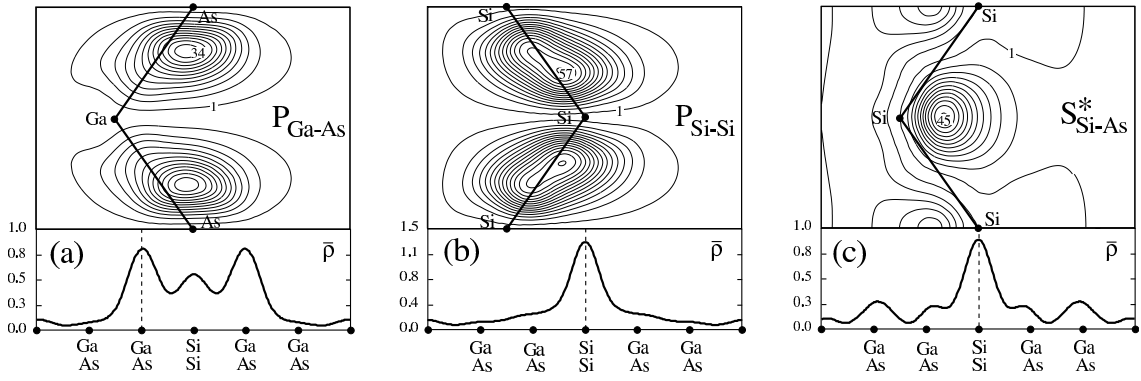}
\caption{\label{state1ML3}
Contour plot of the probability density $\rho$ in the (110) plane indicated by
the dashed line and planar average $(\bar{\rho})$ along the [110] direction of
three states of the GaAs/1\,Si\,ML/GaAs~(110) heterostructure. All states are
at the $\sympt{X}'$ point of the zone. The dashed lines indicate the position
of the planes where the contour plots have been drawn.
}
\end{figure*}

Another state (also symmetric under the $\sigma_{xy}$ reflection) is observed
in the lower part of the GaAs stomach gap, but only near the $\sympt{X}'$ point
(SP$_{\text{Si-As}}$). The latter state is not present in the fully developed
Si/GaAs~(110) heterojunction, and originates mainly from $p$ states of the As
atoms adjacent to the Si layer and $sp$ hybrids of the neighboring Si atoms
(see Fig.~\ref{state1ML2}(a)). The P$_{\text{Si-As}}$ (P$_{\text{Si-Ga}}$)
state derives from the anti-symmetric combination of the P$_{\text{Si-As}}$
(P$_{\text{Si-Ga}}$) interface states on each side of the Si layer (see
Figs.~\ref{state1ML1} and~\ref{stateint1}). The P$_{\text{Ga-As}}$ and
P$_{\text{Si-Si}}$ states are symmetric states, which also derive from their
interface counterparts (see Figs.~\ref{state1ML3} and~\ref{stateint2}).

Finally, we find a well localized conduction state in part of the GaAs
fundamental bandgap (S$^*_{\text{Si-As}}$ in Fig.~\ref{state1ML}) which does
not exist in the fully developed Si/GaAs~(110) heterojunction. The latter state
is an antibonding Si--As state with strong $sp^3$-hybrid character from the Si
atoms occupying the cation sites within the (110) layer (see
Fig.~\ref{state1ML3}(c)). This state is essentially a quantum-well state in its
early stage formation. At larger Si intralayer thickness, the localization of
this state, at specific $\bm{K}$ points of the 2DBZ, will be due to the
conduction band offset of Si in GaAs (i.e., the quantum-well potential, within
an effective-mass description). The S$^*_{\text{Si-As}}$ state is present
within the GaAs bandgap in a large portion of the 2DBZ. Yet, we recall that the
GaAs and Si bandgaps are affected by an error within the LDA-DFT approach,
which may affect also the existence of the S$^*_{\text{Si-As}}$ state. The
calculated GaAs LDA (direct) gap is 0.9~eV, while the experimental
value\cite{lan} is 1.52~eV ($+0.62$~eV correction), and the Si LDA gap is about
0.5~eV to be compared to the experimental value\cite{lan} of 1.17~eV
($+0.67$~eV correction). The gap corrections, however, turn out to be similar
for Si and GaAs; also, the corrections are expected to be rather rigid shifts
throughout the BZ. We therefore do not expect drastic changes in the alignment
of the Si and GaAs conduction bands. To have an estimate of the conduction
offset controlling the localization of the most bound S$^*_{\text{Si-As}}$
states, we should compare the X (indirect) gap of GaAs (its calculated value is
1.4~eV; the experimental value\cite{lan} is 2.18~eV) with the corresponding gap
of Si. Taking into account the 0.14~eV VBO, we thus conclude that the states
near $\sympt{X}'$ are bound by a conduction offset of about 0.87~eV, to be
compared to the theoretical offset of 0.76~eV. This gives us confidence in the
existence of the S$^*_{\text{Si-As}}$ states also in the actual quasiparticle
interface spectrum.

\section{States induced by two Si~(110) monolayers in GaAs}

We end this study of non-polar (110)-oriented structures with the case of 2~MLs
of Si in GaAs. This case is intermediate between the fully developed interface
and the single Si impurity-sheet perturbation. The point group symmetry of the
2~Si~MLs in GaAs contains the reflection operations $\sigma_{yz}$,
$\sigma_{xy}$, and a twofold rotation around the $y$ axis. The point group is
$C_{2v}$, and the space group is non-symmorphic with fractional translation
$(\frac{1}{2},\frac{1}{2},0)$ (in lattice units). Since our calculations do not
take into account the non-scalar spin-orbit interaction, no additional
degeneracy is imposed along the $\sympt{\Gamma}$--$\sympt{X}$ and
$\sympt{\Gamma}$--$\sympt{X}'$ lines by time reversal symmetry. We thus expect
along these lines pairs of symmetric and antisymmetric states for each
atomic-orbital derived state. Conversely, the representations of the groups of
$\sympt{X}$ and $\sympt{X}'$ belong to the two-dimensional irreducible
representation $\sympt{X}_5$, and that of $\sympt{M}$ to the two-dimensional
representation $\sympt{M}_5$ of $C_{2v}$, for which an additional degeneracy is
present.

\begin{figure}[b!]
\includegraphics[width=7cm]{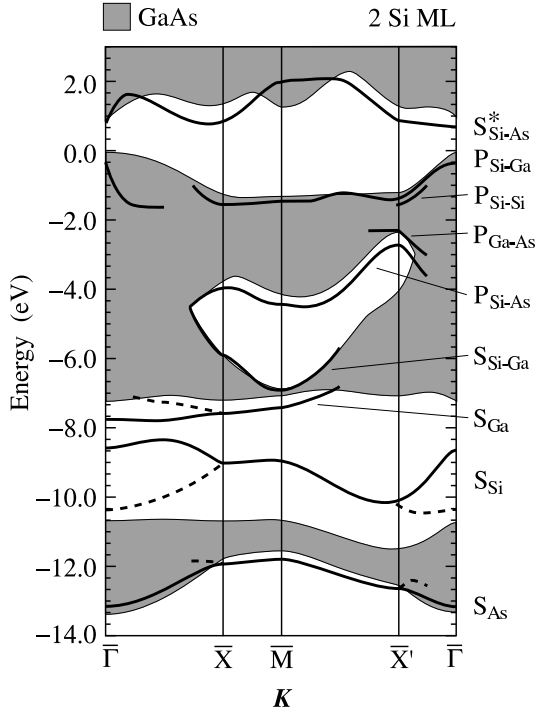}
\caption{\label{state2ML}
Dispersion of the localized and resonant states induced by 2~MLs of Si~(110) in
GaAs relative to the GaAs bulk PBS. Solid lines are the symmetric (by
$\sigma_{xy}$ reflection) states. The dashed-dotted lines represent the
corresponding antisymmetric states.
}
\end{figure}

We plot in Fig.~\ref{state2ML} the states induced by 2~MLs of Si in bulk GaAs.
We have aligned the localized states and resonances with the GaAs PBS only, as
in the 1~ML case. In Fig.~\ref{state2ML}, we distinguished the symmetric (by
$\sigma_{xy}$ reflection) from the antisymmetric states, which have a small
probability density between the adjacent Si layers. Comparing the results for
the 2~Si~MLs with those obtained for 1~Si~ML and the fully developed interface,
we see that the connection between the states of 1~ML and of the interface is
made by the symmetric and antisymmetric states of the 2~MLs structure, which
eventually become degenerate for thicker Si layers. We note that the
S'$_{\text{Ga}}$ and SP$_{\text{Si-As}}$ states are no longer present already
at 2~Si~MLs coverage. Yet, we find a well localized S$^*_{\text{Si-As}}$ state
in a relatively large portion of the 2DBZ. This state is localized in regions
of the 2DBZ, which more closely correspond to the regions exhibiting a large
Si/GaAs conduction band offset, as compared to the 1~ML case. Following the
considerations given for the 1~ML case, this indicates that we are in a
situation in which the quantum well potential is nearly established, with a
potential strength that is becoming similar to the conduction-band offset
between Si and GaAs.

\section{States induced by a Si~(100) bilayer in GaAs}

The situation is more complicated for (100)- than for (110)-oriented
heterovalent heterostructures, because of the polarity of the [100] growth
direction. We mentioned in the Introduction that the fully developed abrupt
Si/GaAs~(100) heterojunction is thermodynamically unstable, as it is
macroscopically charged. Similarly, an abrupt Si~(100) monolayer in GaAs is
also macroscopically charged and thermodynamically unstable. An abrupt Si~(100)
bilayer, instead, gives rise to a dipole, which is the lowest-order multipole
which does not generate a long-range electric field and is hence, in principle,
energetically acceptable. The dipole layer also corresponds to the
highest-order multipole which can induce a change in the band alignment, and is
therefore of possible interest for interfacial band engineering. Here we thus
focus on the bilayer case. Some caution, however, is in order when comparing to
experiment. Although photoemission and internal photoemission studies have
reported some changes in the band alignment in the presence of Si
interlayers,\cite{sorba,titti1,marsi} this effect was found to saturate at Si
coverages of about 0.5~ML, and to decrease in amplitude at higher coverages.
This has been interpreted\cite{mariapert} as an indication of significant
deviations from a Si bilayer distribution above 0.5~ML, due to atomic
interdiffusion. It should be stressed, however, that the positions of the Si
atoms are unknown in any case, and it is only speculatively that one may assume
a dipole layer distribution, even at low coverage. We thus consider the
simplest possible case of 2~MLs of Si in an abrupt bilayer configuration, and
we will end this section with some considerations on the trends one may expect
for bilayers with lower Si nominal concentration. Although 2~MLs of Si in a
bilayer geometry is a structure which has not been produced experimentally so
far, the study of this system is interesting in itself to establish a
relationship between the interface states of the (110) and (100)
heterostructures. We also want to show that intrinsic scattering centers can
exist in the (100) systems, even in an abrupt, defect-free heterostructure.

\begin{figure}[tb]
\includegraphics[width=8.7cm]{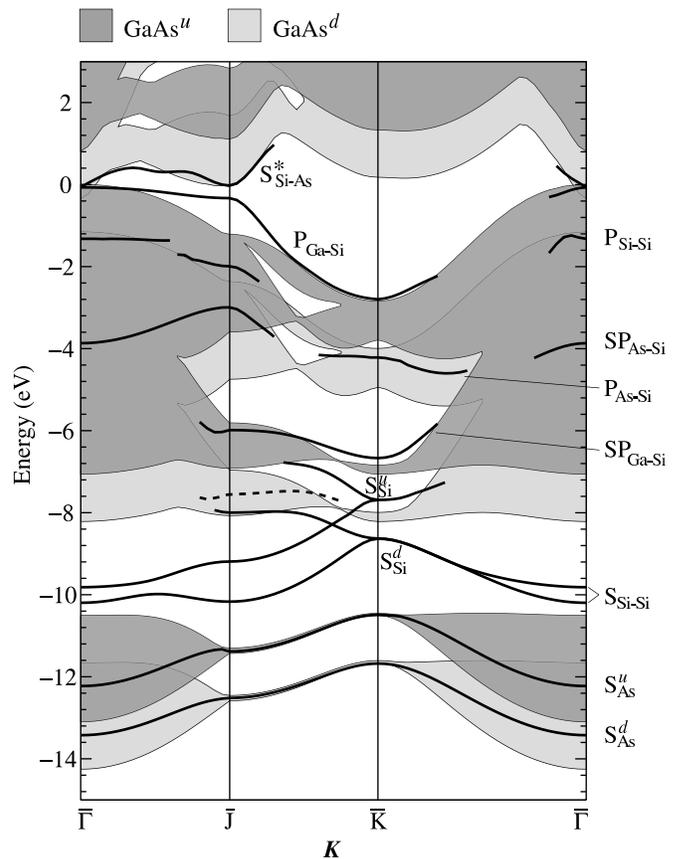}
\caption{\label{stateint2ML100}
Dispersion of the localized and resonant states (thick lines) induced by 2~MLs
of Si in the (100) GaAs homojunction relative to the PBS of the GaAs bulks,
GaAs$^u$ and GaAs$^d$, which are shifted one with respect to the other by the
dipole potential induced by the Si interlayer. The dashed line indicates the
state corresponding to the S$_{\text{Ga}}$ state of the
GaAs/1\,Si\,ML/GaAs~(110) homojunction (see text).
}
\end{figure}

In Fig.~\ref{stateint2ML100}, we show the dispersion of the states induced by
2~MLs of Si in the (100) GaAs homojunction. The PBSs of the two GaAs bulks on
each side of the junction (GaAs$^u$, GaAs$^d$) are rigidly shifted one with
respect to the other by $1.15$~eV, which is the calculated VBO of the relaxed
heterostructure. In the LDA framework, the system is metallic near
$\sympt{\Gamma}$. However, this would not be the case if the experimental gap
of GaAs had been used. We note that the zone center has a very small weight in
the BZ integrations (our grid actually does not include the $\sympt{\Gamma}$
point). The charge transfer generated by the small valence-conduction band
overlap at $\sympt{\Gamma}$ has therefore a negligible influence on the band
alignment (the computed potential lineup $\Delta V$ agrees within 0.1~eV with
the prediction of the linear-response model of Ref.~\onlinecite{mariapert}).
The point group symmetry of the isolated bilayer in GaAs is $C_{2v}$. Since we
have two Si bilayers in the supercell, rotated by $\pi/2$ one with respect to
the other around the twofold $z$ axis, a higher $D_{2d}$ symmetry is present in
the superlattice. The irreducible part of the 2DBZ of our calculation is thus
one half of that of the isolated bilayer, and each point in the irreducible BZ
includes states of both bilayers. However no additional degeneracy is induced
by time reversal symmetry. The bands are shown along the high symmetry lines
connecting $\sympt{\Gamma}=(0,0)$, $\sympt{J}=(\frac{1}{2},\frac{1}{2})$ and
$\sympt{K}=(0,1)$ (in units of $2\pi/a$, where $a$ is the GaAs lattice
constant).

\begin{figure}[tb]
\includegraphics[width=8.7cm]{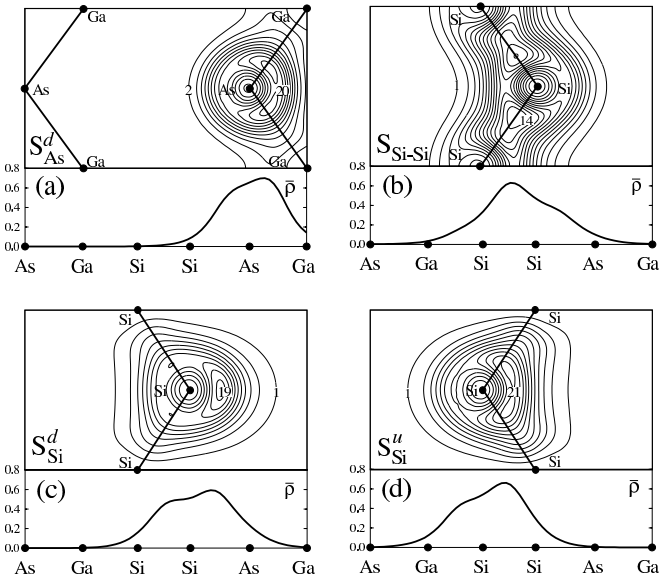}
\caption{\label{state2ML1001}
Contour plot of the probability density $\rho$ in the (110) plane and planar
average $(\bar{\rho})$ along the [100] direction of some states in the
GaAs/2\,Si\,MLs/GaAs~(100) heterostructure. (a) GaAs$^d$ As-derived $s$-state
at $\sympt{J}$. (b) Si--Si-derived $s$-state with lowest energy at $\sympt{J}$.
(c) Si$^d$-derived $s$-state at $\sympt{K}$. (d) Si$^u$-derived $s$-state at
$\sympt{K}$. The probability density for the above states in atomic planes
other than the ones considered is a small fraction of the one shown.
}
\end{figure}

Beginning from the lowest energies, we recover the As-related $s$-state band.
In the present case, pairs of localized $s$ states separated by $\Delta V$
appear, which correspond to As on the upper and lower (S$_{\text{As}}^d$ in
Fig.~\ref{state2ML1001}(a)) energy side of the heterostructure. We observe
three different types of localized Si $s$-like states in the Si~(100) bilayer
structure: one derives mainly from Si atoms sitting on the Ga sites
(S$_{\text{Si}}^d$, see Fig.~\ref{state2ML1001}(c)), the second one is
associated mostly with Si atoms on the As sites (S$_{\text{Si}}^u$, see
Fig.~\ref{state2ML1001}(d)), while the third one involves Si atoms from both
cation and anion sites (S$_{\text{Si-Si}}$ in Fig.~\ref{state2ML1001}(b)), and
shows strong Si--Si bonding features. The first two types of states are found
essentially around the zone edge $\sympt{K}$ point, whereas the
S$_{\text{Si-Si}}$ states are observed in the remaining part of the 2DBZ. The
character of the S$_{\text{Si-Si}}$ states changes progressively along the
$\sympt{J}$--$\sympt{K}$ direction, and matches the character of the
S$_{\text{Si}}^d$ and S$_{\text{Si}}^u$ states at $\sympt{K}$.

\begin{figure}[tb]
\includegraphics[width=7cm]{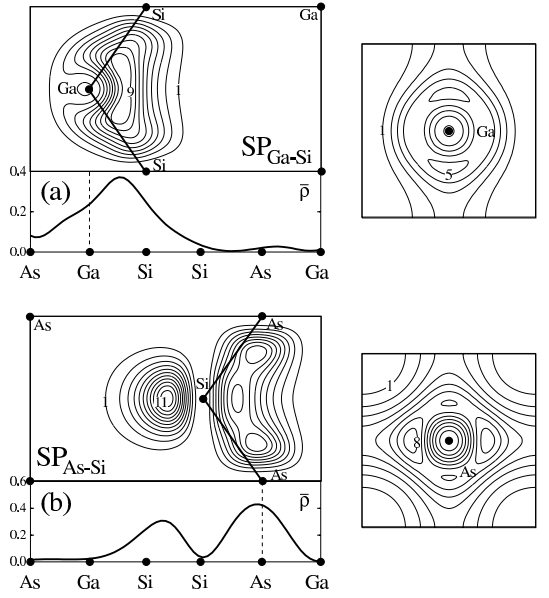}
\caption{\label{state2ML1002}
Left: contour plot of the probability density $\rho$ in the (110) plane and
planar average $(\bar{\rho})$ along the [100] direction of two states in the
(100) GaAs/2\,Si\,MLs/GaAs heterostructure. Right: contour plot of the same
states in the $(100)$ plane indicated by the dashed line. (a) $s$-Ga and $p$-Si
derived state at $\sympt{J}$. (b) $sp$ As and Si-derived state at $\sympt{K}$.
The probability density for the above states in atomic planes other than the
ones considered is a small fraction of the one shown.
}
\end{figure}

\begin{figure}[tb]
\includegraphics[width=8.7cm]{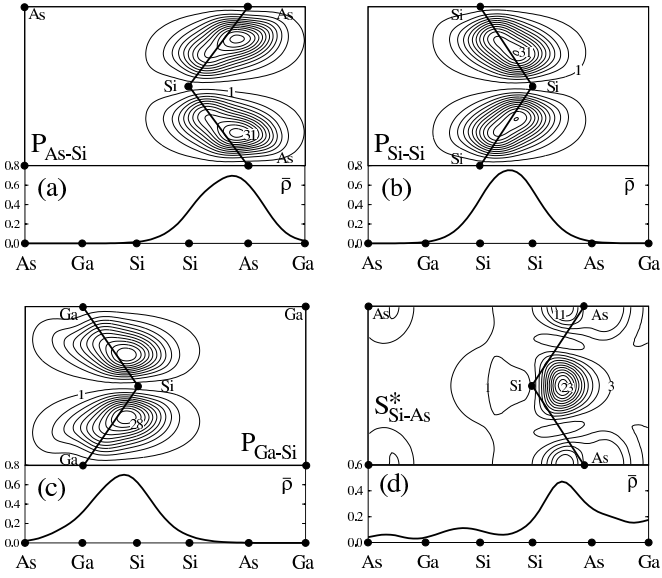}
\caption{\label{state2ML1003}
Contour plot of the probability density $\rho$ in the (110) plane and planar
average $(\bar{\rho})$ along the [100] direction of some states in the (100)
GaAs/2\,Si\,MLs/GaAs heterostructure. (a) GaAs$^d$ As and Si$^d$-derived
$p$-state at $\sympt{J}$. (b) Si-derived $p$-state at $\sympt{J}$. (c) GaAs$^u$
Ga and Si-derived $p$-state at $\sympt{J}$. (d) Si$^d$-derived antibonding
$sp$-state at $\sympt{J}$. The probability density for the above states in
atomic planes other than the ones considered is a small fraction of the one
shown.
}
\end{figure}

The Ga-related localized $s$ state of the (110) GaAs/1\,Si\,ML/GaAs
heterostructure is found as a resonance in the present case (dashed line in
Fig.~\ref{stateint2ML100}). Conversely, we find well-localized states which
derive from interacting Ga $s$ and Si $p$ states (SP$_{\text{Ga-Si}}$ in
Fig.~\ref{state2ML1002}(a)). We also find an SP$_{\text{As-Si}}$ state, which
results from the interaction between As $sp$ hybrids and Si $p$ orbitals
(SP$_{\text{As-Si}}$ in Fig.~\ref{state2ML1002}(b)). The in-plane (110)
P$_{\text{Ga-As}}$ interface state is no longer present, as one might have
expected since the planes of Ga--As zigzag chains are now perpendicular to the
interface. The P$_{\text{Si-Si}}$ state, instead, is found as a bilayer state,
oriented in the (100) direction (see Fig.~\ref{state2ML1003}(b)). We note that
the P$_{\text{Si-Si}}$ state appears only near the valence edge of the GaAs$^d$
PBS, along the $\sympt{\Gamma}$--$\sympt{J}$ line. The state originating from
the interaction between the $p$ states of As and Si (P$_{\text{As-Si}}$ in
Fig.~\ref{state2ML1003}(a)) is found near the zone-edge $\sympt{K}$ point, and
is a resonance in the GaAs$^d$ PBS.

Finally, we also report two states which are present in the main PBS gap (apart
from the metallic region near $\sympt{\Gamma}$). The first one is a Ga--Si
bonding state resulting from the interaction between the $p$ states of Ga and
Si (P$_{\text{Ga-Si}}$, see Fig.~\ref{state2ML1003}(c)) at the GaAs$^u$/Si
junction. The second one is a Si--As antibonding state deriving from $sp^3$
hybrids of the Si and (to a lesser extent) of the As atoms
(S$^*_{\text{Si-As}}$, see Fig.~\ref{state2ML1003}(d)) at the Si/GaAs$^d$
junction. Both states, and in particular the P$_{\text{Ga-Si}}$ state, exhibit
a relatively weak dispersion along the $\sympt{\Gamma}$--$\sympt{J}$ line,
leading to an enhanced DOS near the GaAs$^u$ valence-band edge
(P$_{\text{Ga-Si}}$) and the GaAs$^d$ conduction-band edge
(S$^*_{\text{Si-As}}$).

\begin{figure*}[bt]
\includegraphics[width=15cm]{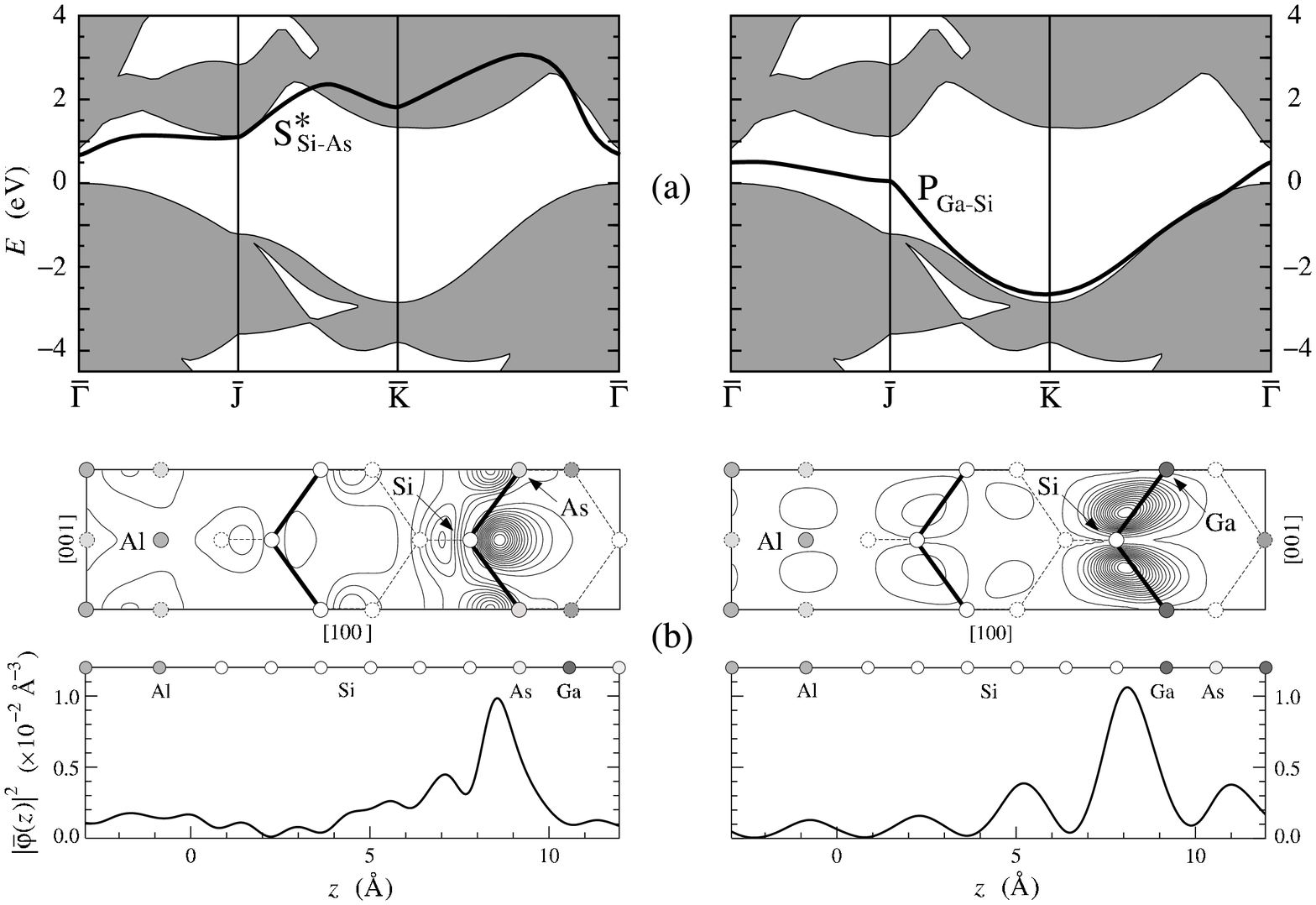}
\caption{\label{resonant}
Resonant Si/GaAs-interface states in the abrupt Al/6\,Si\,MLs/GaAs~(100)
junctions with As- (left) and Ga- (right) terminated GaAs~(100) at the
Si/GaAs~(100) interface. (a) Dispersion of the interface states along the high
symmetry directions of the 2DBZ (solid line); the gray area corresponds to the
projected bulk band structure of GaAs. The zero of energy corresponds to the
GaAs valence band maximum. (b) Probability density of the resonant states at
the $\sympt{\Gamma}$ point. The contour plots are shown in (010) planes
including the As--Si--As (left) and Ga--Si--Ga (right) bonds at the Si/GaAs
interface. Solid (dotted) circles and lines indicate atoms and bonds which are
in (out) of the plane. The planar average of the density is shown below the
contour plots.
}
\end{figure*}

In the presence of only one of the two GaAs$^u$/Si and Si/GaAs$^d$ interfaces,
we expect the resulting P$_{\text{Ga-Si}}$ or S$^*_{\text{Si-As}}$ states to be
{\em within} the GaAs optical gap near $\sympt{\Gamma}$. This is confirmed by
{\it ab initio\/} computations we performed for related engineered systems,
namely abrupt Al/6\,Si\,MLs/GaAs~(100) metal-semiconductor structures with
either an As- or a Ga-GaAs~(100) termination at the polar Si/GaAs
interface.\cite{berthod99} The dispersion of the corresponding
S$^*_{\text{Si-As}}$ and P$_{\text{Ga-Si}}$ states is shown in
Fig.~\ref{resonant}, along the high symmetry direction of the 2DBZ, together
with the probability density of these states. Although the As- and
Ga-terminated Al/6\,Si\,MLs/GaAs~(100) metal/semiconductor structures
incorporate only one of the two types of charged Si/GaAs~(100) interfaces each,
they exhibit no macroscopic electric field in the bulk semiconductor
(GaAs)---because of the presence of the metal---and display either a GaAs$^d$
energy-shifted bulk (in the As-terminated junction) or a GaAs$^u$
energy-shifted bulk (in the Ga-terminated junction). Also, the thickness of the
Si (6 Si monolayers pseudomorphically strained to GaAs) insures negligible
interaction between the continuum of Al bulk states, on one side of the
interlayer, and the resonant S$^*_{\text{Si-As}}$ or P$_{\text{Ga-Si}}$
Si/GaAs-interface states, on the other side of the interlayer. In these
conditions, the As-terminated (Ga-terminated) junction gives rise (see
Fig.~\ref{resonant}) to a band of the S$^*_{\text{Si-As}}$ (P$_{\text{Ga-Si}}$)
Si/GaAs interface states which has a dispersion similar to that of
Fig.~\ref{stateint2ML100} within the GaAs PBSs bandgap, but with an energy at
$\sympt{\Gamma}$ which is well within the gap, namely $\sim 0.2$~eV below
($\sim 0.5$~eV above) the conduction- (valence-) band edge of the GaAs PBS.

The S$^*_{\text{Si-As}}$ and P$_{\text{Ga-Si}}$ states that we find at the
abrupt As-terminated and Ga-terminated Si/GaAs~(100) interfaces, respectively,
are the (100) analogues of the S$^*_{\text{Si-As}}$ and P$_{\text{Si-Ga}}$
states that we observed in the non-polar GaAs/Si/GaAs~(110) heterostructures.
However, contrary to the case of the globally neutral Si/GaAs~(110) junction,
at the As- terminated Si/GaAs~(100) interface, the attractive potential created
by the globally positive excess ionic charge of the As--Si donor bonds is
sufficiently strong to pull down S$^*_{\text{Si-As}}$ conduction states within
the GaAs fundamental gap. Similarly, the potential created by the globally
negative ionic charge of the Ga--Si acceptor bonds is able to push up
P$_{\text{Ga-Si}}$ states from the valence band into the GaAs gap. Hence, the
S$^*_{\text{Si-As}}$ (P$_{\text{Ga-Si}}$) states become potential scattering
centers at the polar As-terminated (Ga-terminated) Si/GaAs~(100)
heterojunction.

As mentioned before, Si may be confined in a bilayer geometry only at nominal
coverages lower than or equal to 0.5~ML. We have therefore also investigated
what happens to the S$^*_{\text{Si-As}}$ and P$_{\text{Ga-Si}}$ states when the
Si coverage is reduced to 0.5~ML. We have considered two different ordered
GaAs/Si/GaAs ($2\times2$) interlayer atomic configurations corresponding each
to one Si $\rightarrow$ Ga and one Si $\rightarrow$ As atomic substitution per
($2\times2$) surface-unit cell within the GaAs bilayer. In the first
configuration, the pair of Si atoms sit on adjacent Ga-As sites, so that each
Si atom has three As or three Ga nearest neighbors and one Si nearest neighbor
(configuration A), whereas in the second configuration each Si atom has four As
or four Ga nearest neighbors, and no Si nearest neighbor (configuration
B).\cite{noteAB}

In both A and B configurations, we find that the S$^*_{\text{Si-As}}$
(P$_{\text{Ga-Si}}$) states associated with the Si--As (Si--Ga) donor
(acceptor) bonds become resonances at $\sympt{\Gamma}$. The energy positions of
the resonances depend somewhat on the Si arrangement. The S$^*_{\text{Si-As}}$
resonance is located at $+0.2$~eV from the conduction band edge in the A
configuration, and at $+0.1$~eV from the same edge in the B configuration. The
P$_{\text{Ga-Si}}$ resonance remains, instead, a near-band edge resonance
(within 0.1~eV from the valence band edge) in both cases.
We note that, in both cases, the conduction resonance gives rise to a stronger
DOS feature than the valence resonance, due to the lower density of bulk states
present near the conduction edge. The conduction resonance may be responsible
for a peculiar feature observed in internal photoemission experiments by
dell'Orto {\it et al.}\cite{titti} The associated increased DOS at
$\sympt{\Gamma}$ is indeed detectable, in principle, by transport
measurements.\cite{titti}

\section{Koster-Slater model for the interface-state problem}

The {\it ab initio\/} calculations of the preceding sections have provided us
with the energy position and atomic-like character of various states localized
at the Si/GaAs interface. However, it is not simple to extract from the
full-fledged {\it ab initio\/} calculations a general understanding of the
mechanisms of interface-state formation. For example, it is evident from
Fig.~\ref{stateint} that except for the S$_{\text{As}}$ states which form a
complete band along the high-symmetry lines, the other states appear only along
specific lines or near specific points of the 2DBZ. This is a common feature
observed for many different interfaces studied in the literature. In this
section, we thus try to rationalize the above results based on a more general,
yet simplified description, that contains the main physical ingredients of the
problem. Specifically, one would like such a model to include the effects of
({\it i\/}) the strength and sign of the local potential at the interface
generated by the different chemical nature of the interface atomic
constituents, ({\it ii\/}) the band structures of the two bulk materials, and
({\it iii\/}) the macroscopic lineup of the two bulk band structures across the
interface. We report below a simple analytical criterion for the existence of
interface states, which explicitly relates the existence of these states to
({\it i\/}), ({\it ii\/}), and ({\it iii\/}). A derivation of the analytical
form of this criterion, which is mathematically simple and physically
transparent, is also presented. A detailed and rigorous quantum-mechanical
derivation, together with the explicit expressions for the microscopic
interface-specific terms, is given in Ref.~\onlinecite{max}.

An interface can be regarded as a perturbation of the bonding structures of two
different bulk materials.\cite{pollman} Also, since for stable interfaces the
perturbation involved is neutral and short ranged, we may formulate the problem
in terms of few perturbed layers, and thus derive a criterion similar to the
one Koster and Slater introduced for deep impurities in
semiconductors.\cite{KoSl54} Following the authors of
Ref.~\onlinecite{pollman}, we construct the interface as a perturbation of two
bulk semiconductors $A$ and $B$. We write the unperturbed Hamiltonian as
	\begin{equation}
		H_0 = \left(\begin{array}{cc}
			H^A_0 & 0\\0 & H^B_0
		\end{array}\right),
	\end{equation}
where $H^A_0$ and $H^B_0$ are the Hamiltonians of the two bulk periodic
crystals, with standard periodic boundary conditions over $N$ layers. The
corresponding one-particle Green's function of the system is
	\begin{equation}
		G_0 = \left(\begin{array}{cc}
		G^A_0 & 0\\0 & G^B_0
		\end{array} \right),
	\end{equation}
with $G^A_0$ and $G^B_0$ the Green's functions of the Hamiltonians $H^A_0$ and
$H^B_0$, respectively. A perturbation $U$ (which will be specified below) is
then introduced; $U$ acts on this system, and in particular couples $A$ and
$B$, and yields an interface $A/B$ (plus two free surfaces). The total
Hamiltonian describing the perturbed system is $H=H_0+U$. A basis set of
Wannier functions is assumed in the following discussion. Specifically, as $U$
preserves the periodicity in planes parallel to the interface, the basis
functions at a given $\bm{K}$-point in the 2DBZ consists of localized-layer
Wannier functions\cite{pollman} in $A$ and $B$, obtained from the Bloch
functions of the corresponding bulk with wave-vector projection in the 2DBZ
equal to $\bm{K}$.

\begin{figure}[tb]
\includegraphics[width=8.7cm]{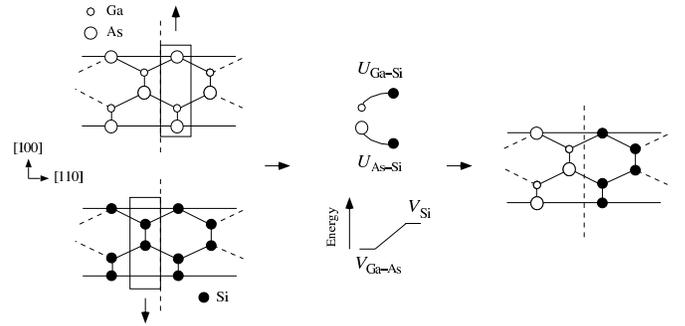}
\caption{\label{figintschema}
Schematic description of the interface creation procedure discussed in the
text.
}
\end{figure}

The construction of the interface, and thus of the perturbation $U$, is
schematized in Fig.~\ref{figintschema}. We consider the Si/GaAs~(110)
interface, and assume for simplicity that the interactions are described by
tight-binding nearest-neighbor Hamiltonians. To keep the expressions as simple
as possible, we also refer in the following to one band
$\varepsilon_{A(B)}(\bm{K},k_z)$ for each bulk semiconductor. The
generalization to the multiband case is straightforward.\cite{max} A layer of
atoms of material $A$ (e.g., GaAs) and a layer of atoms $B$ (Si) are first
removed on each side of the interface in bulk space $A$ and $B$, respectively.
This procedure is analogous to the creation of free surfaces and can be
performed in practice by changing the on-site parameters $V_A$ and $V_B$, which
eventually are let go to infinity.\cite{pollman} Then, the interactions between
the atoms of the left-hand side $A$ and right-hand side $B$ surfaces (Si--Ga
and Si--As) are switched on across the interface. These interactions are
represented by off-diagonal terms $U_{AB}$, which are nearest-neighbor
tight-binding matrix elements coupling the two interface layers. Finally, the
charge transfer at the interface, i.e., the perturbation induced by the
interface relative to a superposition of bulk crystal potentials, is added as
on-site diagonal matrix elements $U_A$ and $U_B$ for the semiconductor
interface layer $A$ and $B$, respectively. Since these perturbations are
measured with respect to the average electrostatic potential of the
corresponding bulks, the lineup is introduced in the Green's functions of the
bulks. If we assume that the removal procedure described above has been
performed, the only relevant block of the matrix $U$ is
	\begin{equation}
		U = \left(\begin{array}{cc}
		U_A & U_{AB}\\U^{\dagger}_{AB} & U_B
		\end{array} \right),
	\end{equation}
where all elements are scalars in the one-band model. The condition for having
bound states can then be derived from the determinantal equation\cite{economou}
	\begin{equation}\label{eq:det}
		\det[\openone - G_0 (E)U]=0,
	\end{equation}
and reads
	\begin{equation}\label{eqcondint}
		1-U_A G^A_0-U_B G^B_0-(|U_{AB}|^2-U_A U_B)G^A_0 G^B_0 = 0.
	\end{equation}
If the zero of energy is fixed at the top of the valence band
$\varepsilon_A(\bm{K},k_z)$, the Green's functions are:
	\begin{equation}
		G^A_0=\frac{1}{N}\sum_{k_z}\frac{1}{E-\varepsilon_A(\bm{K},k_z)},
	\end{equation}
and
	\begin{equation}
		G^B_0=\frac{1}{N}\sum_{k_z}
			\frac{1}{E-\varepsilon_B(\bm{K},k_z)-\text{VBO}(A/B)}.
	\end{equation}
The condition (\ref{eqcondint}) tends to the usual Koster-Slater relation, $1-
UG_0= 0$, for deep impurities ($U$ being the potential perturbation induced by
the impurity in the host described by $G_0$) when only one semiconductor and
on-site interactions are retained. Given the linear relation, $ G_0(E) = \int
d\varepsilon D(\varepsilon) /(E-\varepsilon)$, between the Green's function of
a bulk and the corresponding electronic density of states (DOS),
$D(\varepsilon)$, we find from Eq. (\ref{eqcondint}) that, for a given set of
parameters ($U_A$, $U_B$ and $U_{AB}$), the existence of bound states and
resonances depends on: ({\it i\/}) the strength of the DOS of each bulk band
with respect to the corresponding on-site parameter $U_{A(B)}^{-1}$, and ({\it
ii\/}) the amplitude of the {\em product} of the DOS's of the two bulk bands
with respect to the interface coupling parameter $|U_{AB}|^{-2}$. Localized
states may appear above/below the bulk bands according not only to the strength
of the on-site terms $U_A$ and $U_B$ at the interface, but also to the {\em
relative strength\/} of these terms and the coupling terms $U_{AB}$.\cite{max}
In fact, when $U_A=U_B=0$, $U_{AB}$ can push states both above {\em and} below
the bulk bands. It should be noted, however, that when the band offset is
non-vanishing, even in a one-dimensional model, the on-site perturbations
generated by the interface may induce resonances rather than bound states. This
is in contrast to the case of deep impurities, where a state is always bound in
one dimension (for local potentials).

\begin{figure*}[tb]
\includegraphics[width=14cm]{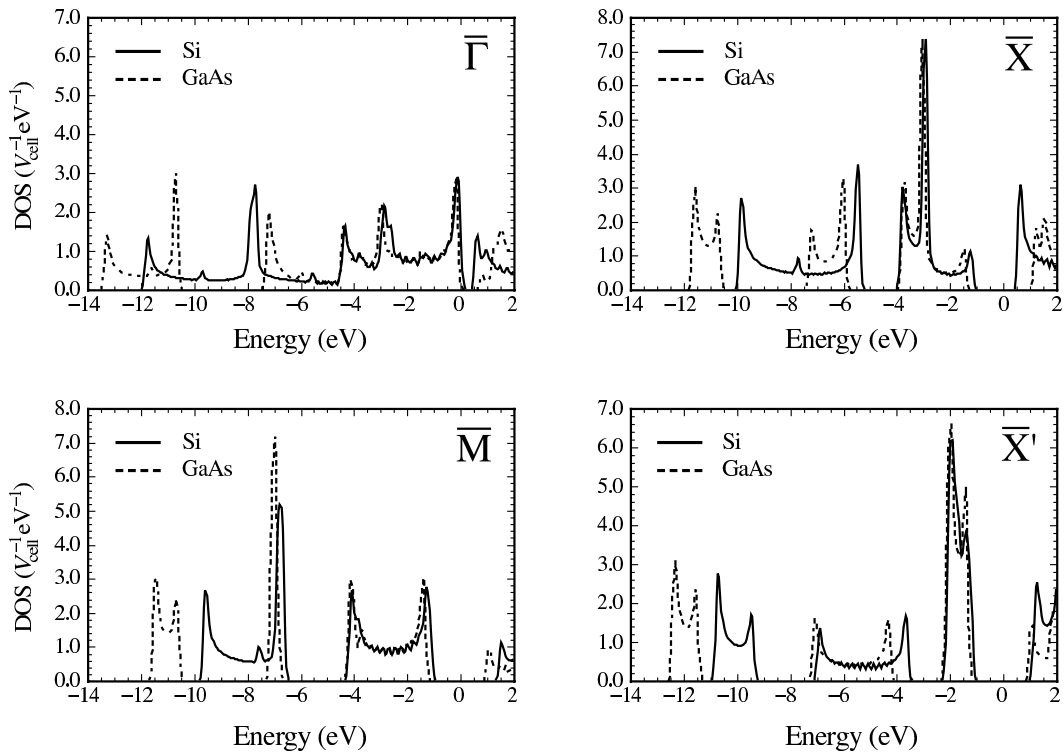}
\caption{\label{statehs}
Wave-vector resolved density of states (DOS) of bulk Si and GaAs for the four
high-symmetry points of the (110) 2DBZ, and integrated along a line in
$k$-space parallel to the [110] crystallographic direction. The two DOS are
aligned using the calculated VBO of 0.14~eV at the Si/GaAs~(110) interface. The
volume of normalization is the GaAs unit-cell volume. The degeneracies of the
Si (GaAs) valence-band DOS features, given in order of increasing energy, are:
1 (1, 1) at $\sympt{\Gamma}$, 2, 2 (1, 1, 2) at $\sympt{X}$ and $\sympt{M}$,
and 1, 1, 2 (1, 1, 2) at $\sympt{X}'$.
}
\end{figure*}

With the help of Eq. (\ref{eqcondint}), we can now try to identify the regions
and points of the 2DBZ where interface states are most likely to appear. In
view of the relationship between Green's function and DOS, we have examined the
DOS of both Si and GaAs bulks, for a given point of the (110) 2DBZ, integrated
along a line in $k$-space parallel to the growth direction. The result is
plotted in Fig.~\ref{statehs} for the four high-symmetry points of the (110)
2DBZ. The DOS has been calculated using tetragonal (properly strained)
supercells of 4 atoms oriented along the (110) direction for both bulks. For
each point in the 2DBZ we sampled the (110) direction with 26 $k$-points
distributed on a uniform grid. A Gaussian broadening of 0.1~eV was used. We
have employed the same procedure as in the previous section to align the DOS's
of the two bulks.

The $s$ DOS of GaAs is a sharp feature, well separated from the Si bands at
$\sympt{X}$, $\sympt{M}$, and $\sympt{X}'$. The localized S$_{\text{As}}$ state
is induced thus by an on-site perturbation. The $U_{\text{As}}$ on-site
potential is attractive,\cite{max} due to the attractive potential of the
nearest Si atoms, and is responsible for the existence of this state. The
S$_{\text{Si-Ga}}$ states originate mainly from both the Si and GaAs peaks
located at about $-6$~eV and $-7$~eV in the $\sympt{X}$ and $\sympt{M}$ DOS's
spectra, respectively. These edges are associated with Si (GaAs) electronic
states that contain an important $s$ component, together with a non-negligible
$p$ contribution, from the Si (Ga) atoms. The on-site $U_{\text{Ga}}$ and
$U_{\text{Si}}$ matrix elements of adjacent Ga and Si atoms are
repulsive.\cite{max} However, contrary to the case of the As $s$ states, the Si
and GaAs edges are separated by less than 1~eV, and the $U_{AB}$ term also
contributes to the formation of the S$_{\text{Si-Ga}}$ states. At
$\sympt{\Gamma}$ and $\sympt{X}'$, instead, no comparable DOS features are
detected at similar energies.

At $\sympt{X}'$, two pronounced and strongly overlapping Si and GaAs DOS
features are visible: with band edges at about $-2.5$~eV and $-1$~eV. The lower
edges contribute predominantly to the P$_{\text{Si-As}}$ and P$_{\text{Ga-As}}$
states, and the upper ones to the P$_{\text{Si-Si}}$ and P$_{\text{Si-Ga}}$
states. However, only the P$_{\text{Si-As}}$ and P$_{\text{Si-Ga}}$ states can
be followed along the 2DBZ up to the $\sympt{X}$ point. The strong overlap near
the band edges make the contribution of the $U_{AB}$ term important, as
evidenced by the mixed Si-GaAs nature of the resulting interface states. These
states are localized by the attractive (repulsive) ionic potential in the
Si--As (Si--Ga) bonding region.\cite{max} At the $\sympt{\Gamma}$ point, no
bound state is present in the fundamental gap. This is due to the attractive
character of the $U_{\rm As}$ and $U_{\rm Si}$ matrix elements,\cite{max} and
to the finite value of the band offset. The off-diagonal term $U_{AB}$ has thus
not enough strength to push a state in the gap; we note that the situation
changes, instead, at the polar, negatively charged Ga-terminated Si/GaAs~(100)
interface, where the repulsive potential induced by the negative charge pushes
the P$_{\text{Si-Ga}}$ state within the GaAs bandgap. Similar considerations
apply to other abrupt interfaces studied in the
literature,\cite{baraf,pickett,pollman,saito,wang,laref,apl2001} giving us
confidence in the soundness of the above model description. We also note that
estimates for the interface-bonding parameters may be obtained from relatively
simple models, such as the tight-binding approach of Ref.~\onlinecite{pollman}.

\section{summary and conclusions}

Using an {\it ab initio\/} pseudopotential approach, we have studied the states
induced by thin Si interlayers in abrupt GaAs/Si/GaAs~(110) and (100)
heterostructures. We have also investigated the interface states of the fully
developed Si/GaAs~(110) heterojunction and resonant Si/GaAs~(100) interface
states occurring within the GaAs bandgap in Al/6\,Si\,MLs/GaAs~(100)
heterojunctions. We have examined the bonding properties and atomic character
of these various states, and their possible connections.

The results reveal interesting common features between localized states and
resonances in structures with different interface orientations and, for a given
orientation, between states localized by 1 or 2 Si monolayers and the interface
states of the fully developed Si/GaAs junction. By studying the evolution of
the resonances and localized states with Si-interlayer thickness, we were able
to explain the origin of Si-related resonant interface states occurring in the
fully developed Si/GaAs~(110) junction. The {\it ab initio\/} results also
indicated that most of the localized states of the Si/GaAs~(110) interface
persist up to 2 and even 1 monolayer Si-interlayer coverage. Moreover, the most
striking valence- (conduction-) near-band edge states found in the
GaAs/2\,Si\,MLs/GaAs~(100) heterostructure were also shown to originate from
localized As--Si donor-bond (Ga--Si acceptor-bond) states of the fully
developed As-terminated (Ga-terminated) Si/GaAs~(100) interface.

In order to gain further insight into the mechanism of formation of interface
states, we have presented a Koster-Slater type model developed for the
interface-state problem. A simple analytical criterion for the existence of
interface states [Eq.~(\ref{eqcondint})] has been derived, which allows us to
predict the existence and/or energy of interface states from the DOS structure
of the constituent materials and from some interface bonding
parameters.\cite{max} With this model, we were able to explain the different
type of localized states obtained from the {\it ab initio\/} calculations for
the GaAs/Si interface, and account for their energy position and location
within the Brillouin zone. More generally, this model is expected to be useful
to predict trends due to changes in the interface atomic structure (e.g., due
to atomic intermixing\cite{max}) or in the electronic band structure of the
bulk materials (e.g., due to alloying or different polytypes\cite{apl2001}).

\begin{acknowledgments}

We acknowledge the hospitality and support of the Ecole Polytechnique
F{\'e}d{\'e}rale of Lausanne where most of this work has been performed while
two of us (MD and CB) were completing their PhD thesis.

\end{acknowledgments}


\begin{thebibliography}{99}
\bibitem{notesym}
We note that localized interface states may exist also within the bulk PBS if
their symmetry differs from that of the bulk states. This is, however,
relatively unlikely in semiconductor heterojunctions, especially at generic $k$
points of the BZ.
\bibitem{noteAB}
Configuration A is found to be lower in energy by $\sim 0.2$~eV per Si atom
than configuration B.
\end{thebibliography}
\end{document}